\documentclass{article}

\usepackage{PRIMEarxiv}

\usepackage[utf8]{inputenc} 
\usepackage[T1]{fontenc}    
\usepackage{hyperref}       
\usepackage{url}            
\usepackage{booktabs}       
\usepackage{amsfonts}       
\usepackage{nicefrac}       
\usepackage{microtype}      
\usepackage{lipsum}
\usepackage{fancyhdr}       
\usepackage{graphicx}       
\graphicspath{{media/}}     
\usepackage{cite}
\usepackage{amsmath,amssymb,amsfonts}
\usepackage{algorithmic}
\usepackage{graphicx}
\usepackage{textcomp}
\usepackage[ruled,linesnumbered]{algorithm2e}
\usepackage{bm}
\usepackage{pifont}

\usepackage[OT1]{fontenc}
\usepackage{booktabs}
\usepackage[table,xcdraw]{xcolor}
\usepackage{multirow}
\usepackage{adjustbox}
\usepackage{bm}
\usepackage[labelformat=simple]{subcaption}
\usepackage[misc]{ifsym}

\title{Segmentation and Vascular Vectorization for Coronary Artery by Geometry-based Cascaded Neural Network
\thanks{This paper has been submitted to TMI.} 
}

\author{Xiaoyu~Yang \textsuperscript{1,2}, 
Lijian~Xu \textsuperscript{2,3 ({\Letter}) },
Simon~Yu \textsuperscript{4},
Qing~Xia \textsuperscript{5},
Hongsheng~Li \textsuperscript{6}\&
Shaoting~Zhang \textsuperscript{2}\\
\textsuperscript{1} College of Electronics and Information Engineering, Tongji University \\
\textsuperscript{2} Shanghai Artificial Intelligence Laboratory \\
\textsuperscript{3} Centre for Perceptual and Interactive Intelligence, the Chinese University of Hong Kong \\
\textsuperscript{4} Department of Imaging and Interventional Radiology, the Chinese University of Hong Kong \\
\textsuperscript{5} SenseTime Research \\
\textsuperscript{6} Department of Electronic Engineering, the Chinese University of Hong Kong
}

\begin{document}
\maketitle

\begin{abstract}

  Segmentation of the coronary artery is an important task for the quantitative analysis of coronary computed tomography angiography (CCTA) images and is being stimulated by the field of deep learning. However, the complex structures with tiny and narrow branches of the coronary artery bring it a great challenge. Coupled with the medical image limitations of low resolution and poor contrast, fragmentations of segmented vessels frequently occur in the prediction. Therefore, a geometry-based cascaded segmentation method is proposed for the coronary artery, which has the following innovations: 1) Integrating geometric deformation networks, we design a cascaded network for segmenting the coronary artery and vectorizing results. The generated meshes of the coronary artery are continuous and accurate for twisted and sophisticated coronary artery structures, without fragmentations. 2) Different from mesh annotations generated by the traditional marching cube method from voxel-based labels, a finer vectorized mesh of the coronary artery is reconstructed with the regularized morphology. The novel mesh annotation benefits the geometry-based segmentation network, avoiding bifurcation adhesion and point cloud dispersion in intricate branches. 3) A dataset named CCA-200 is collected, consisting of 200 CCTA images with coronary artery disease. The ground truths of 200 cases are coronary internal diameter annotations by professional radiologists. Extensive experiments verify our method on our collected dataset CCA-200 and public ASOCA dataset, with a Dice of 0.778 on CCA-200 and 0.895 on ASOCA, showing superior results. Especially, our geometry-based model generates an accurate, intact and smooth coronary artery, devoid of any fragmentations of segmented vessels.

\end{abstract}

\keywords{Segmentation \and Coronary Artery \and Geometry-based \and Mesh Annotation}

\section{Introduction}
\label{sec:introduction}

Knowledge of the coronary artery anatomy is a prerequisite for many clinical applications. The segmentation and vascular vectorization in coronary computed tomography angiography (CCTA) images can be very valuable for the analysis of the anatomy and functions of the coronary artery. With the modeling of the coronary artery, doctors can quickly and accurately locate, assess and diagnose plaques and stenoses in the blood vessels. Beyond diagnosis, coronary segmentation can also inform the navigation and planning of cardiac interventions by determining the optimal catheter path, stent location and size, among other information, which can improve the safety and efficiency of the procedure. In this context, the automatic segmentation of the coronary artery is of great importance in clinics.

However, automating the segmentation of coronary artery remains arduous. The coronary artery has a unique tree structure with thin and narrow branches that vary greatly. Distal branches are too slender to be segmented precisely, especially when other blood vessels interfere. Moreover, the sparsity and anisotropy of CCTA
images result in most segmentation methods being voxel-based. The reconstructed mesh from the
voxel-based segmentation mask is rough with a noticeable lattice shape. Furthermore, CCTA images have limitations such as low resolution and poor contrast, which make the coronary artery segmentation more difficult.

Currently, deep learning methods have been widely employed in the coronary artery segmentation \cite{songAutomaticCoronaryArtery2022,leeTeTrISTemplateTransformer2019,zhuSegmentationCoronaryArteries2022,zhangProgressiveDeepSegmentation2022,panCoronaryArterySegmentation2021,shenCoronaryArteriesSegmentation2019,tianAutomaticCoronaryArtery2021,zhuCoronaryAngiographyImage2021,huangCoronaryArterySegmentation2018}, which mainly generated voxel-based masks based on the Unet architecture. Nonetheless, automatic segmentation that preserves the integrity and continuity of the coronary artery remains challenging due to the common fragmentation of the segmented vessels. Meanwhile, the mesh-deformation-based methods have been increasingly drawing the attention of the community. Nevertheless, they only focus on large and regular organs, such as the liver and hippocampus. The coronary artery with its intricate structures and narrow branches is hard to go directly from voxel-based segmented results to mesh.

To tackle the aforementioned problems, a new workflow is firstly designed from  voxel-based coronary artery labels for producing realistic annotation of vectorized mesh. Subsequently, the generated vectorized meshes are utilized as training annotations in the following neural networks, providing a more accurate, intact and smooth morphology of the coronary artery, especially at the stenosis regions. Furthermore, a novel cascaded mesh segmentation network is presented, where the generated vectorized coronary artery mesh becomes more integrated compared to the voxel-based segmentation results. Finally, the coronary artery mesh results are smoother with plentiful details, particularly in tiny and narrow branches. Point cloud from the coronary artery mesh is capable of being directly used in the diagnosis, skipping the step of reconstruction from voxel-based segmentation results. Finally, extensive experiments demonstrate the robustness and feasibility of our method. 

\section{Related Work}

\subsection{Coronary Artery Segmentation}

Traditional methods for coronary artery segmentation are mainly divided into two categories: region growing \cite{stutzmannAutomaticSegmentationAorta2010,bouraoui3DSegmentationCoronary2010,wangAutomaticSegmentationCoronary2012,cetinVesselTractographyUsing2013,tejero-de-pablosBeamStackSearchBased2021,chenAutomaticSegmentationCoronary2012a,gharleghiAutomatedCoronaryArtery2022} and partitioning methods \cite{duAutomatedCoronaryArtery2021a,lugauerPreciseLumenSegmentation2014}. Region growing performs iteratively adding similar neighboring voxels so that each final region encompasses a single class. It mainly includes level sets methods \cite{stutzmannAutomaticSegmentationAorta2010, wangAutomaticSegmentationCoronary2012}, snake models \cite{bouraoui3DSegmentationCoronary2010} and tracking method \cite{cetinVesselTractographyUsing2013, tejero-de-pablosBeamStackSearchBased2021}. Whereas, region growing relies on several flexible parameters, which are difficult to be determined in specific cases. Partitioning methods implement grouping regions with similar properties together including preserving the coronary artery as a separate region. The main method used for partitioning is clustering, where the Hessian matrix is usually assisted to enhance the image. But, the segmentation results of the coronary artery are not precise, absent smoothness and details of the shape.

Recently, deep learning has shown its feasibility of coronary artery segmentation with excellent performance, surpassing traditional algorithms in terms of accuracy. Meanwhile, most of the current methods \cite{songAutomaticCoronaryArtery2022,leeTeTrISTemplateTransformer2019,zhuSegmentationCoronaryArteries2022,zhangProgressiveDeepSegmentation2022,panCoronaryArterySegmentation2021,shenCoronaryArteriesSegmentation2019,tianAutomaticCoronaryArtery2021,zhuCoronaryAngiographyImage2021,huangCoronaryArterySegmentation2018} perform voxel-based segmentation and achieve improvements based on the Unet. 3D-FFR-Unet \cite{songAutomaticCoronaryArtery2022} proposes integrating the dense convolutional block to achieve effective feature extraction and fusion, improving the segmentation accuracy of the coronary artery. TETRIS \cite{leeTeTrISTemplateTransformer2019} proposes a template transformer network to improve the segmentation  performance of the coronary artery, where a shape template is deformed to match the underlying structure of interest through a trained spatial transformer network. FFNet \cite{zhuSegmentationCoronaryArteries2022} fuses spatio-temporal features, which are extracted by the Unet, to improve the segmentation results. PDS \cite{zhangProgressiveDeepSegmentation2022} achieves coronary artery segmentation by leveraging contextual anatomical information and vascular topologies through their proposed SAD module and HTL module. TreeConvGRU \cite{kongLearningTreestructuredRepresentation2020} designs the 
tree-structured convolutional gated recurrent unit (ConvGRU) model to learn the anatomical structure of the coronary artery.

Besides, the centerline exhibits a crucial facilitator in the segmentation of the coronary artery. Along centerlines, GCN predicts the radii to obtain the coronary artery mesh\cite{gaoJointCoronaryCenterline2021, wolterinkGraphConvolutionalNetworks2019}. Similarly, WHD \cite{huangCoronaryWallSegmentation2020} uses the centerline to separately segment the inner lumen and outer vessel wall with contour-regularized weighted Hausdorff distance loss. TreeConvGRU\cite{kongLearningTreestructuredRepresentation2020} traverses the entire coronary artery tree through the centerline.

\subsection{Mesh Segmentation Network}

Instead of traditional voxel-based segmentation, more studies are concentrating on integrating the mesh deformation neural network into segmentation tasks. SAN \cite{yaoIntegrating3DGeometry2019} explicitly incorporates 3D geometry into classical 3D FCNs for better liver segmentation. The 3D point cloud is projected from voxel-based extracted image features and deformed via a GCN-based shape-aware network for segmentation. Similarly, Voxel2Mesh \cite{wickramasingheVoxel2Mesh3DMesh2020a} extends pixel2mesh \cite{wangPixel2MeshGenerating3D2018} to 3D images for segmentation tasks of the liver, synaptic junction, and hippocampus. MSMR \cite{zhaoSegmentationTrueLumen2022a} applies mesh segmentation in the lumen of aortic dissection (AD), which has an explicit tubular structure. AD morphology constrains the initial mesh and guides the deformation, which improves the efficiency of the deep network and avoids down-sampling. GMB \cite{wangGeometricMorphologyBased2021} exploits point net to refine voxel-based coronary artery segmentation results by removing irrelevant vessels, where point cloud and voxel-based segmentation results are converted into each other. However, current methods of integrating mesh deformation networks are limited to big organs with regular shapes, such as the liver. The coronary artery has an explicit tree structure, with tiny and narrow branches, that current graph neural networks are hard to perform such complex mesh deformations. It is a great challenge to achieve vectorial segmentation of the coronary artery.

\section{Methodology}

In this section, a new method of generating elaborate mesh annotation is firstly introduced for geometrical regularization of the coronary artery segmentation. Then, we concentrate on the proposed geometry-based cascaded segmentation for the coronary artery.

\subsection{Fine Mesh Annotation for Geometrical Regularization}

\begin{algorithm}[htbp]
	\caption{Framework of Generating Elaborate Mesh Annotation for Coronary Artery.}
	\label{alg:mesh_annotation}
	\KwIn{Voxel Annotation of the Coronary Artery $L$.}
    \KwOut{Mesh Annotation of the Coronary Artery $M$.}  
	\BlankLine

    Obtain key points $P$ of the coronary artery by skeletonizing the voxel annotation $L$;

    Acquire key points $P_{K}$ of each branch $K$ by splitting the coronary artery.

    \ForEach{Coronary Artery Branch $K$}{

        Simulate the centreline of the coronary artery branch through key points $P_{K}$ by B-spline;
    
        \ForEach{$P_{Ki}$}{
            Compute the Tangential direction of the key point $P_{Ki}$;

            Sample rays at the cross-section of the key point $P_{Ki}$;        

            Calculate the intersection between each ray and voxel annotation $L$ of the coronary artery boundary.

            Smooth the radius from the key point $P_{Ki}$ to the intersection.

            Generate the cross-sectional boundary $M_{Ki}$ of the vectorized mesh of the coronary artery.
        }

        Derive the vectorized mesh of each branch $M_{K}$ by connecting adjacent cross-sectional boundaries $M_{Ki}$.

        Smooth the vectorized mesh of each branch $M_{K}$ along the centerline $P_{K}$.        
    }

    Generate the complete vectorized mesh annotation $M$ of the coronary artery by merging each branch $M_{K}$.
	
\end{algorithm}

The framework of our method for generating the fine mesh annotation of the coronary artery is shown in Algorithm.\ref{alg:mesh_annotation}, consisting of three main processes: skeletonization, reconstruction and integration. Through skeletonization, key points of the coronary artery tree are extracted and split into individual branches. Using these key points and coronary artery annotation, each branch is reconstructed with a smooth surface. While dealing with the intricate multi-forks of the coronary artery, individual branches are integrated to form a more realistic vessel shape. The above steps will be described in detail.

In terms of skeletonization, the Deep Reinforced Tree-Traversal Agent (DRT) \cite{liDeepReinforcedTreeTraversal2021} is employed to extract the key points of the coronary artery and establish the tree structure, which is our preliminary work. Considered the key points of the coronary artery tree, the line connecting the head point and each branch endpoint of the coronary artery is considered a centerline of the coronary artery branch. The centerline of each branch is interpolated with a cubic B-spline curve, and key points are sampled at every 0.2 mm.

\begin{figure}[htbp]
    \centering
    \includegraphics[width=0.45\textwidth]{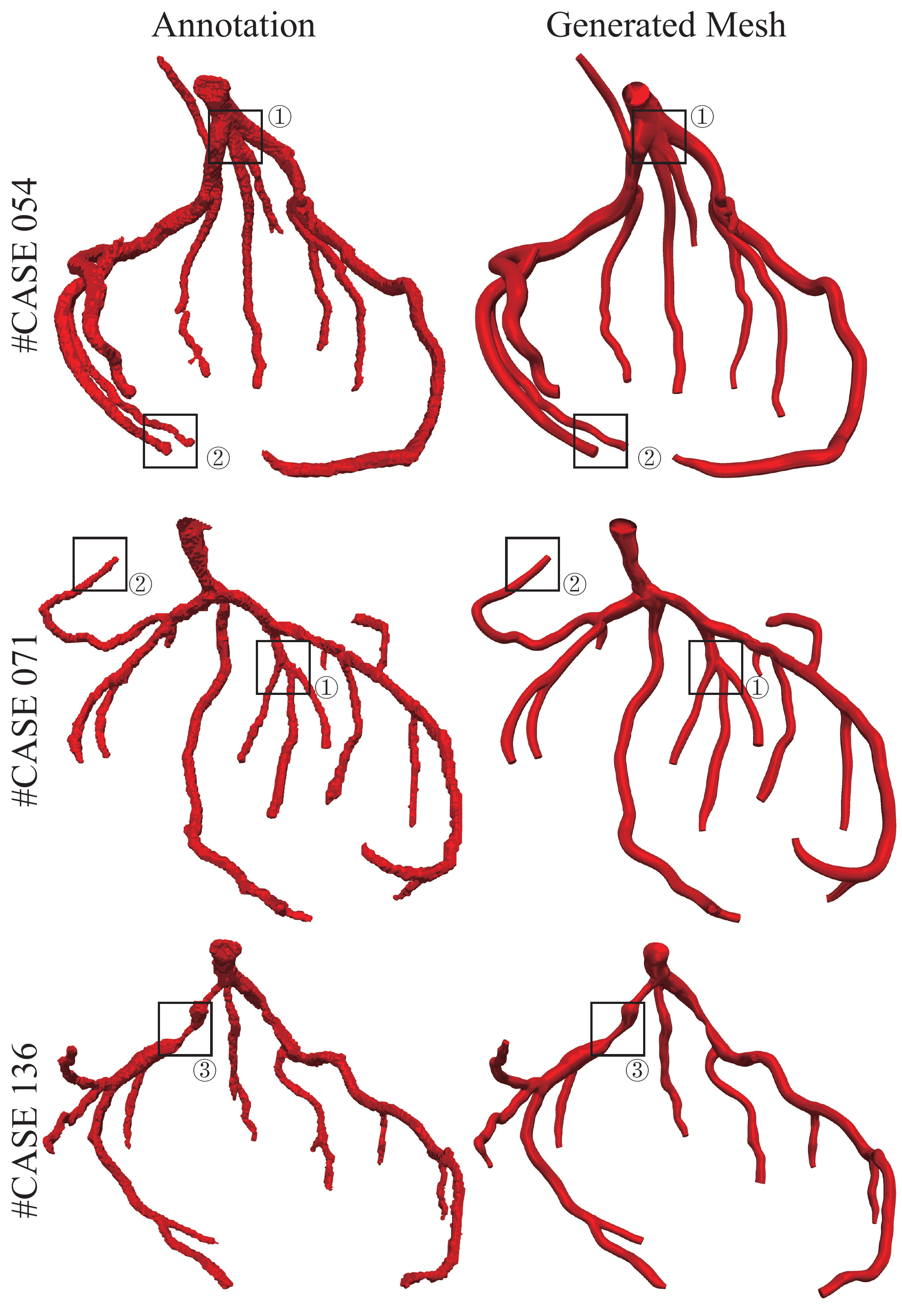}
    \caption{Results of generated vectorized mesh annotation. \ding{172}:Multi-forks of coronary \ding{173}:Tiny and narrow branches \ding{174}: Stenotic vessels compressed by plaque.}
    \label{fig:generated-mesh-annotation}
\end{figure}

Then, reconstruction is applied with the key points of each coronary artery branch. The tangent of the key point is calculated and served as the normal vector to form the cross-section of the coronary artery. At each cross-section, rays are sampled at every $15^{\circ}$ in a counterclockwise direction from the key points and intersect with the voxel annotation to form a mesh layer of the coronary artery boundary. However, since the sparsity of the voxel annotation, which consists of discrete voxels, the sampled boundary of the coronary artery in the cross-section is rough and not entirely smooth. In order to restore the original morphology of the coronary artery as much as feasible, $\text{1-d}$ gaussian filter is applied to smooth the radii from keypoint $P_{Ki}$ to every boundary point $A^{j}_{Ki}$, where $j$ denotes the angle of the ray. Following the formation of smooth coronary artery borders in each cross-section, boundary points of two adjacent cross-sections form triangular patches to compose the coronary artery mesh. Besides the smoothness of the cross-sectional boundary, the vectorized mesh of the coronary artery needs to be flattened along the centerline, demonstrating the context smoothness.

Finally, with the coronary artery mesh for each branch, the mesh boolean union operation is implemented to merge them and finish the complete coronary artery mesh. Unlike prior methods, we reconstruct each branch mesh and merge them separately, rather than generating the entire coronary artery mesh together. It avoids intricate modeling and massive computation of the coronary artery forks, particularly in trifurcation. Furthermore, since each branch of the bifurcation contains the same trunk, the transition is smoother and closer to the real coronary artery vessel.

\begin{figure*}[htbp]
  \centering
  \centering
  \includegraphics[width=0.98\textwidth]{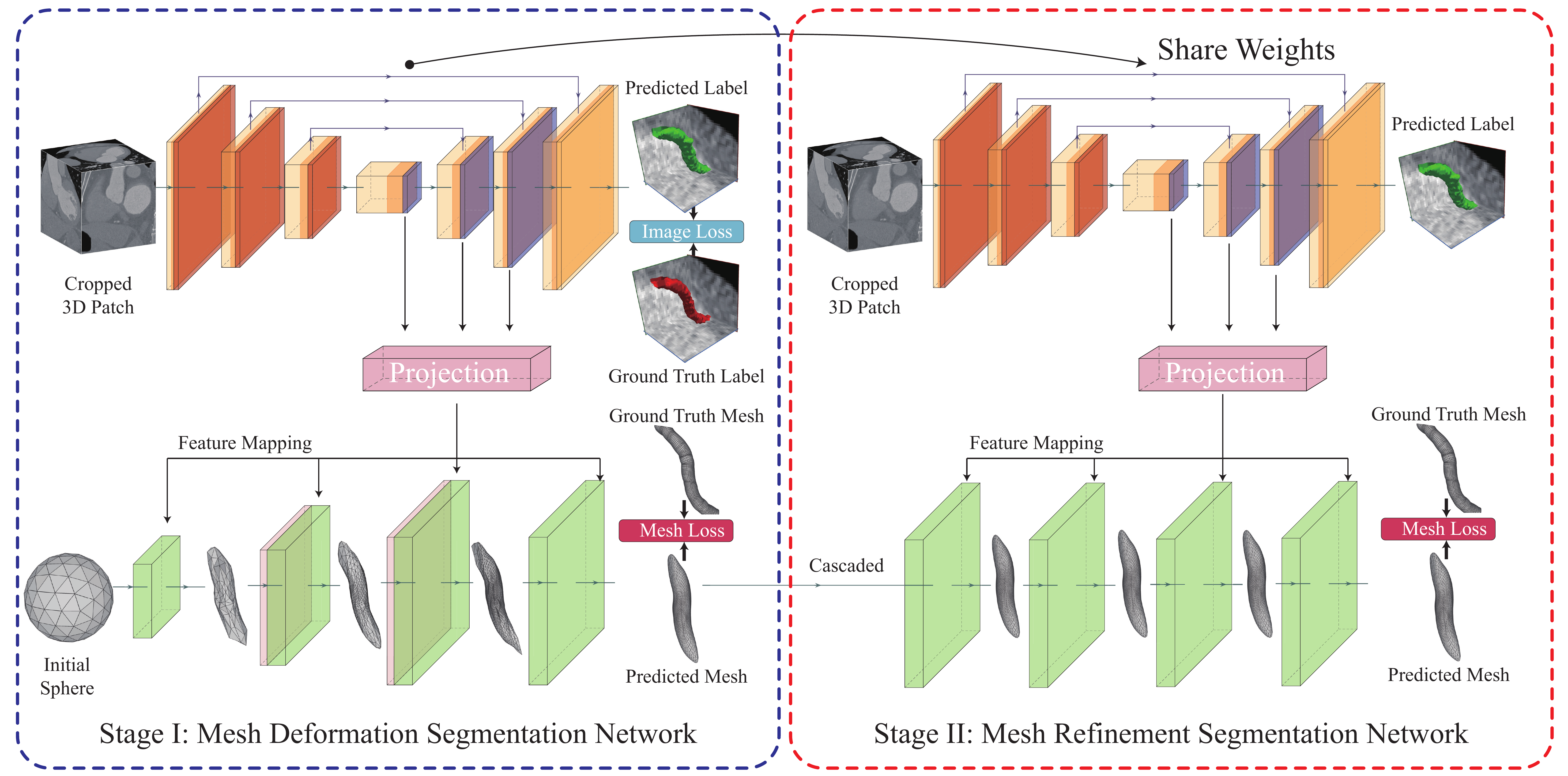}
  \caption{Our geometry-based cascaded segmentation network for generating mesh of the coronary artery.}
  \label{fig:workflow-e2e}
\end{figure*}

Generated vectorized mesh annotations are shown in Fig.\ref{fig:generated-mesh-annotation}. The left is reconstructed results from voxel-based segmentation labels using the marching cube method, and the right denotes generated mesh masks by our algorithm. Our approach is capable of reconstructing the smooth coronary artery surface, with abundant details of tiny and narrow branches. Moreover, it is generalized to cope with various complex coronary artery structures, such as trifurcation, and even four-forks. The transition at the junction of the multi-forks is natural and realistic. For vessels that are compressed at the plaques, our reconstructions are also closer to reality, conserving the tubular morphology of the coronary artery.

\subsection{Geometry-based Cascaded Segmentation Network}

Aiming at generating the vectorized mesh of the coronary artery directly, an geometry-based cascaded neural network is presented as shown in Fig.\ref{fig:workflow-e2e}, consisting of two steps: mesh deformation and refinement.

At stage I, given a cropped 3D patch of the CCTA images $\mathbf{X}\in \mathbb{R}^{L\times H\times W}$, a classical U-shape neural network is trained to extract image features of the coronary artery under the framework of voxel-based segmentation. Guided by the projected image features of the U-shape network, a graph convolutional network (GCN) is applied to deform the mesh, achieving the vectorization of the segmentation results. The U-shape network and the GCN are trained together. At stage II, the previous U-shape network is fixed and applied to extract image features without training. The coarse mesh of the coronary artery is input into a new GCN without unpooling, cascading the two steps and generating the fine mesh of the coronary artery. The details are as follows.

\textbf{Graph Convolutional Network}: a sphere mesh $\mathcal{G}=\{\mathcal{V},\mathcal{E}\}$  with 162 vertices and 480 edges is initialized as the input of the GCN, where $\mathcal{V}$ denotes the set of vertices and $\mathcal{E}$ represents the set of edges. The mesh with $N$ vertices $v_{i}\in\mathcal{V}$ in the GCN has its adjacency matrix $\mathbf{A}\in \mathbb{R}^{N\times N}$ and diagonal degree $\mathbf{\hat{D}}_{ii} = \sum_{j=0} \mathbf{\hat{A}}_{ij}$, where $\mathbf{\hat{A}} = \mathbf{A} + \mathbf{I}$. The graph convolution is executed as Eq.\ref{Eq.1}.
\begin{equation}
    \label{Eq.1}
    \mathbf{V}^{\prime} = \mathbf{\hat{D}}^{-1/2} \mathbf{\hat{A}} \mathbf{\hat{D}}^{-1/2} \mathbf{V} \mathbf{\Theta}    
\end{equation}
where $\mathbf{\Theta}$ represents the parameters of the neural network and $\mathbf{V}\in\mathbb{R}^{N\times C}$ symbolizes the feature vector with $C$-dimension for each node $v_{i}$. In addition, the residual block is applied to predict the deformation of the mesh instead of predicting the vertices location of the target mesh directly, which simplifies the difficulty of training. Furthermore, the initial sphere is easily deformed but lacks enough details of the coronary artery. Graph unpooling is implemented in our GCN at stage I, dividing one triangular face into four parts along the midpoint of each side and assigning the mean feature vector of one edge to the node of the midpoint. It supplements more vertices and edges, retouching the mesh of the coronary artery. The LNS \cite{wickramasingheVoxel2Mesh3DMesh2020a} strategy is performed to project extracted image features into the mesh space.

\textbf{Optimization of Segmentation Network}: For jointly training the U-shape neural networks and GCN, various loss functions are adopted to optimize them. First, image loss is mainly driving the U-shape network under the voxel-based segmentation framework, consisting of SoftDice loss and cross-entropy loss. Second, mesh loss optimizes the GCN, including chamfer distance loss, laplacian smoothing, normal consistency loss and edge loss. The chamfer distance dominates the optimization of the GCN, which measures the distance of two point clouds between the prediction and ground truth as Eq.\ref{Eq.chamfer}, guiding the deformation of the mesh. 
\begin{equation}
    \label{Eq.chamfer}
    \begin{aligned}
        \mathcal{L}_{CD}\left(\mathcal{V}_1, \mathcal{V}_2\right)
        &=\frac{1}{\left|\mathcal{V}_1\right|} \sum_{x \in \mathcal{V}_1} \min _{y \in \mathcal{V}_2}\|x-y\|_2^2 \\
        &+\frac{1}{\left|\mathcal{V}_2\right|} \sum_{y \in \mathcal{V}_2} \min _{x \in \mathcal{V}_1}\|x-y\|_2^2            
    \end{aligned}
\end{equation}
Laplacian smoothing (Lap) and normal consistency loss (NC) are utilized to regularize the smoothness of the mesh. Laplacian smoothing $\mathcal{L}_{Lap}$ computes the uniform weights of all edges connected at a vertex. Normal consistency loss computes the angle of the normal $n_{0}$ and $n_{1}$ for each pair of neighboring faces as Eq.\ref{Eq.NL}.
\begin{equation}
    \label{Eq.NL}
    \mathcal{L}_{NC} = \sum_{e\in \mathcal{E}} {1 - \cos{(n_{0},n_{1})}}
\end{equation}
Besides, edge loss $\mathcal{L}_{EG}$ computes the length of each edge, avoiding outlier vertices. In summary, the total loss of the GCN is shown in Eq.\ref{Eq.gcn}.
\begin{equation}
    \label{Eq.gcn}
    \mathcal{L}_{GCN} = \lambda_{1}\mathcal{L}_{CD} + \lambda_{2}\mathcal{L}_{Lap} + \lambda_{3}\mathcal{L}_{NC} + \lambda_{4}\mathcal{L}_{EG}
\end{equation}
where $\lambda_{1-4}$ represents the weight of each loss.

\textbf{Regularization of The GCN Training}: The intricate structure of the coronary artery presents a great challenge for the neural network. GCN is hard to learn such complicated morphology. Therefore, cropped coronary artery mesh is classified into two categories: tube and bifurcation. Compared with twisted, irregular and multi-forks coronary artery mesh, tube and bifurcation have simpler morphology, which is more straightforward to be learned by the neural network. Hence, morphological regularization is presented to regularize cropped mesh into tube or bifurcation. Through morphological regularization, the geometry-based neural networks can learn more precisely the geometrical features of the coronary artery.

\section{Experiments}

In this section, the datasets and evaluation metrics are first introduced. Then, the improvement brought by vectorized mesh annotation is validated through the ablation experiments. Finally, two datasets are used to extensively demonstrate the robustness and feasibility of coronary artery segmentation results generated by our model.

\subsection{Implementation Details}

Our proposed method is evaluated on a public coronary artery dataset ASOCA and a collected dataset CCA-200.

\textbf{1. ASOCA} \cite{gharleghi2022automated,gharleghi2022computed}: ASOCA dataset contains 40 training cases and 20 testing cases, and 30 of these patients report having coronary artery disease. The collected images have an anisotropic resolution, with an in-plane resolution of 0.3-0.4 mm and out of plane resolution of 0.625 mm.

\textbf{2. CCA-200}: 200 cases with coronary artery disease are collected named CCA-200 dataset. To demonstrate the robustness of our model in small-scale data, comparative experiments are designed: 20 cases are used for training, and 180 cases for testing. The collected images are acquired with an isotropic resolution of 0.5 mm. Ground truths of 200 cases are coronary artery internal diameter annotations labeled by four radiologists.

The evaluation consists of various metrics, including Dice, Hausdorff distance (HD), average symmetric surface distance(ASSD), chamfer distance (CD), Smooth and our proposed Num of Segments (NoS). Dice assesses the overlap between the predicted results and the ground truth. HD, ASSD and CD measure the geometrical morphology of the generated results. Smooth is determined by calculating the normal consistency of the adjacent faces in the reconstructed mesh, revealing the smoothness and flatness of the results. Furthermore, to highlight the fragmentation problem encountered by voxel/pixel-based methods on the segmentation of the coronary artery, the metric Num of Segments (NoS) is proposed to count the number of connecting vessels for assessing the integrity and continuity of the coronary artery. We run all the experiments on NVIDIA A100 (80GB) GPU, Pytorch 2.0. The Adam is used to optimize the network with the initial learning rate of 0.001. 

\begin{figure}[htbp]
  \centering
  \includegraphics[width=0.48\textwidth]{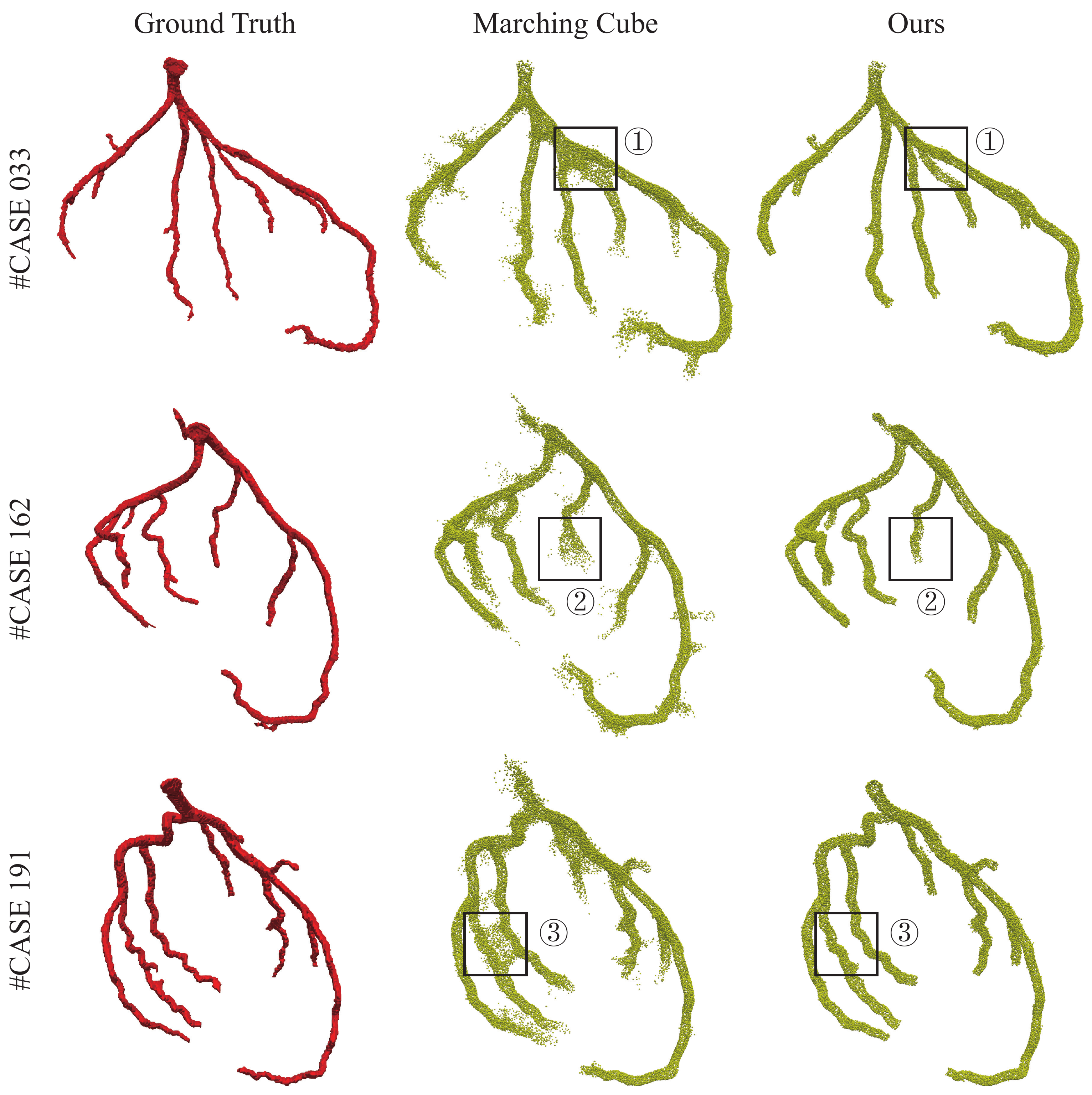}
  \caption{Comparsion Results of Ablation Experiments.\ding{172}: Multi-forks of coronary \ding{173}: Tiny and narrow branches \ding{174}: closely adjacent branches}
  \label{fig:ablation}
\end{figure}


\subsection{Ablation Experiments}

Ablation experiments are performed to demonstrate the feasibility and improvement of generated vectorized mesh annotation for the geometry-based coronary artery segmentation network in our collected dataset. For comparison, the traditional marching cube method is utilized to produce mesh annotations, which usually appear in other geometry-based segmentation methods, such as liver segmentation. The generated coronary artery point clouds are exhibited in Fig.\ref{fig:ablation}, clearly presenting the internal structure of the predicted mesh. Using our refined centerline-based annotation, the geometry-based segmentation network is qualified to outline the boundary of the coronary artery, absent dispersion of the point cloud. As shown at \ding{172}, the trifurcation of the coronary artery in Fig.\ref{fig:ablation}, a clear and natural intersection is formed devoid of the diffusion of points. As for tiny and narrow branch ends of \ding{173}, the geometry-based segmentation network trained by our vectorized mesh annotation will not induce points dispersion as marching cube annotations do. Besides, the adhesion effect is particularly pronounced at closely adjacent branches of \ding{174} using marching cube mesh annotations, whereas segmentation results trained by our centerline-based annotations can clearly maintain the morphology of each branch. In CCTA images with low image resolution and poor contrast, our centerline-based approach can still generate a refined mesh with compact and clear coronary artery boundaries, whereas the marching cube will synthesize blurred and sticky mesh annotation at multi-forks and tiny branches, especially in complicated coronary artery structures.

Moreover, quantitative evaluation verifies the effect of our centerline-based mesh annotation on improving geometry-based coronary artery segmentation. Points in generated coronary artery point cloud less than 0.5 mm from the voxel-based coronary artery annotation are considered hits, and vice versa are considered misclassified. By counting the number of hits, the point cloud hit ratio is calculated, that the higher the more accurate the coronary artery segmented point cloud. Compared with the traditional marching cubes (MC) methods for generating coarse mesh annotation, our model achieves a precision of 0.96, a recall of 0.85, an F1 of 0.88 and an accuracy of 0.85, surpassing the MC method with a precision of 0.92, a recall of 0.85, an F1 of 0.8 and an accuracy of 0.76. It presents that our fine centerline-based mesh annotations can significantly improve the segmentation results for the complicated coronary artery.

\subsection{Overall Evaluation}

In this part, three main representative types of coronary artery segmentation methods are conducted to compare comprehensive experiments on our collected CCA-200 dataset and ASOCA dataset, which are 2D pixel-based, 3D voxel-based and geometry-based segmentation methods, respectively.

\begin{figure*}[htbp]
  \centering
  \includegraphics[width=0.98\textwidth]{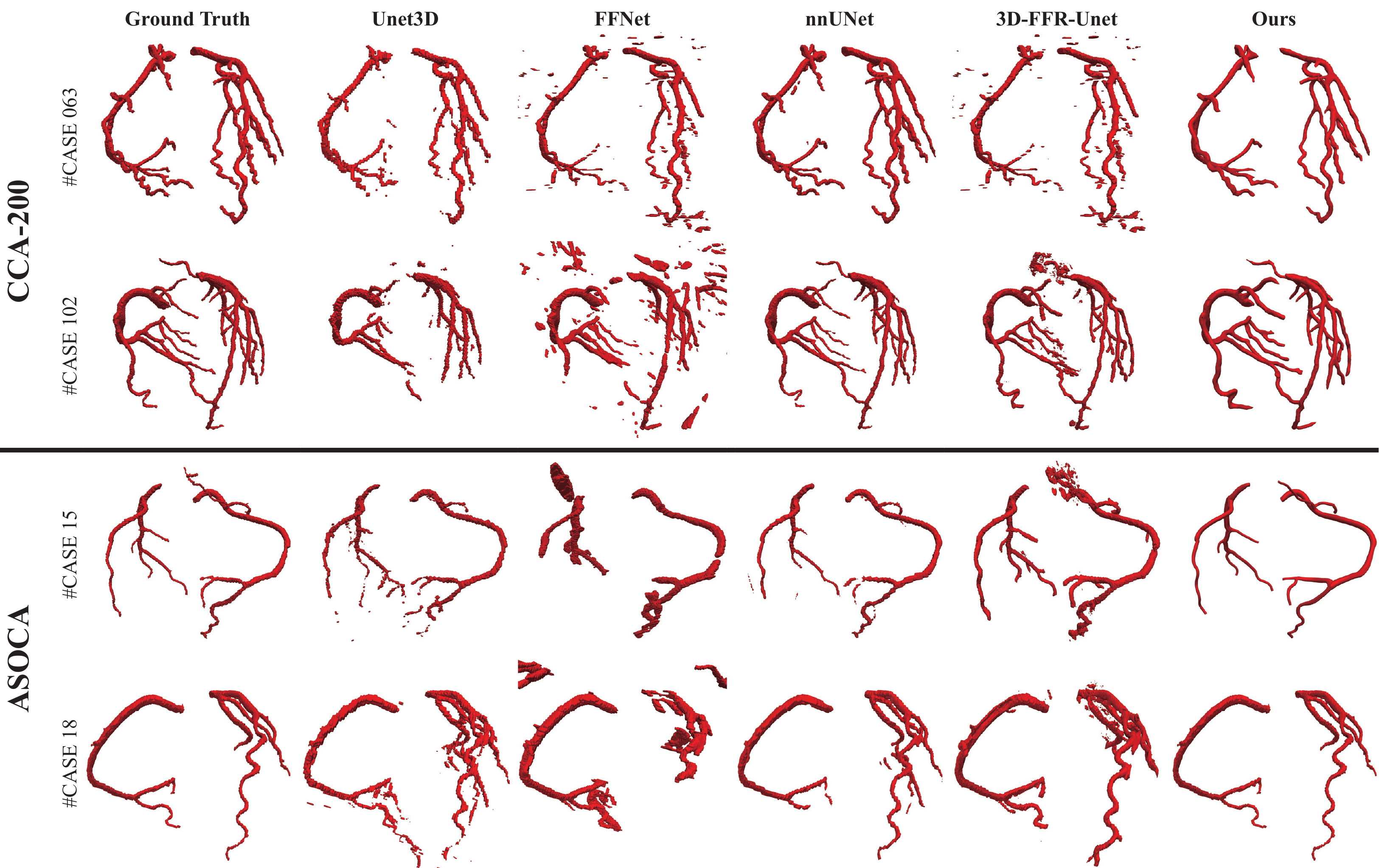}
  \caption{Comparsion Results of Comprehensive Experiments on our collected CCA-200 dataset and public ASOCA dataset with current mainstream methods. Compared with the ASOCA, the coronary artery in our collected CCA-200 has more complicated and elaborate structures, including more multi-forks, and more twisted and narrower branches, which raises higher demand for the segmentation of the coronary artery.}
  \label{fig:exp}
\end{figure*}

Intuitively, Fig.\ref{fig:exp} presents the coronary artery segmentation results of different methods on our collected CCA-200 dataset and public ASOCA dataset, respectively. It can be seen in Fig.\ref{fig:exp}, that fragmentations of segmented vessels frequently occur in voxel-based segmentation, especially for the coronary artery with complicated and twisted structures, such as our collected CCA-200. Conversely, our geometry-based method preserves the complete and elaborate coronary artery, elegantly avoids the fragmentations attributing to the geometry-based segmentation network. Moreover, the tiny and narrow branches of the coronary artery are more accurately and precisely delineated, eliminating the limitations of sparsity and the low resolution of CCTA images. Besides, with the vectorization, the overall segmentation results of the coronary artery conserve the smoothness of the vessel, compenstating for more realistic morphology.

\begin{table}[htbp]
    \renewcommand\arraystretch{1.2}
    \setlength{\tabcolsep}{1.8mm}{
        \begin{tabular}{@{}lcccccc@{}}
        \toprule
                    & Types                                                                                                                     & Dice           & HD            & NoS           & Smooth        & CD            \\ \midrule
        ResUnet       \cite{zhangRoadExtractionDeep2018}            & \multirow{2}{*}{\begin{tabular}[c]{@{}c@{}}2D Pixel\\ based\end{tabular}} & 0.579          & 3.79          & 110.6         & 0.76          & 105.88        \\
        H-DenseUnet   \cite{liHDenseUNetHybridDensely2018}          &                                                                           & 0.586          & 6.08          & 117.3         & 0.79          & 195.47        \\ \midrule
        Unet3D        \cite{cicek3DUNetLearning2016}                & \multirow{4}{*}{\begin{tabular}[c]{@{}c@{}}3D Voxel\\ based\end{tabular}} & 0.641          & 3.39          & 61.8          & 0.63          & 68.11         \\
        nnUnet        \cite{isenseeNnUNetSelfconfiguringMethod2021} &                                                                           & 0.753          & 1.83          & 12.9          & 0.79          & 34.90         \\
        FFNet         \cite{zhuSegmentationCoronaryArteries2022}    &                                                                           & 0.685          & 3.26          & 100.0         & 0.74          & 59.80         \\
        3D-FFR-Unet   \cite{songAutomaticCoronaryArtery2022}        &                                                                           & 0.758          & 0.84          & 161.6         & 0.81          & 7.08          \\ \midrule
        Voxel2Mesh    \cite{wickramasingheVoxel2Mesh3DMesh2020a}    & \multirow{2}{*}{\begin{tabular}[c]{@{}c@{}}Geometry\\ based\end{tabular}} & 0.191          & 28.86         & 2.0           & 0.06          & 519.61        \\
        \textbf{Ours}                                               &                                                                           & \textbf{0.778} & \textbf{0.31} & \textbf{2.0}  & \textbf{0.05} & \textbf{2.57} \\ \bottomrule
        \end{tabular}
    }
    \caption{Quantitative Evaluation Results of the Coronary Artery Segmentation for Different Methods on CCA-200 Dataset.}
    \label{table:CCA-200}  
\end{table}

Quantitatively, the evaluation results of the coronary artery segmentation for different methods on CCA-200 dataset are presented on Table.\ref{table:CCA-200}. The complicated and elaborate structures of the coronary artery on our CCA-200 dataset allow for a comprehensive assessment of the method's performance in sophisticated and realistic scenarios. From the aspect of overlap with the ground truth, our method achieves the Dice of 0.778, surpassing other methods. In term of the geometrical morphology, the HD of 0.31 and the chamfer distance of 2.57 are smaller than others, indicating a more similar morphology to the realistic coronary artery. Particularly, voxel/pixel-based methods inevitably produce fragmentations of the segmented vessels in their predicted results, exhibiting a high NoS. Whereas, the mesh deformation of our geometry-based segmentation network guarantees the continuous integrity of the coronary artery results with NoS of 2. Left and right coronary arteries are produced completely. Voxel2Mesh also generates only 2 parts of the coronary artery, but it cannot cope with the mesh deformation from one initial sphere into multiple  complex branches of the coronary artery, resulting in a particularly low Dice. In addition, benefiting from the vectorization of our geometry-based segmentation network, the coronary artery results of our model have a more smooth and flat surface with a Smooth of 0.05, overcoming the limitations of sparsity and low resolution of the CCTA images.

\begin{table}[htbp]
  \renewcommand\arraystretch{1.2}
  \setlength{\tabcolsep}{3mm}{
      \begin{tabular}{@{}lccccc@{}}
          \toprule
                                                                    & Dice           & HD             & \multicolumn{1}{l}{ASSD}          & \multicolumn{1}{l}{Smooth}          & \multicolumn{1}{l}{NoS} \\ \midrule
          ResUnet     \cite{zhangRoadExtractionDeep2018}            & 0.780          & 0.723          & 1.13                              & 0.605                               & 49.2                    \\
          H-DenseUnet \cite{liHDenseUNetHybridDensely2018}          & 0.853          & 0.388          & 0.73                              & 0.785                               & 36.7                    \\
          Unet3D      \cite{cicek3DUNetLearning2016}                & 0.846          & 0.395          & 0.72                              & 0.633                               & 36.5                    \\
          nnUnet      \cite{isenseeNnUNetSelfconfiguringMethod2021} & 0.859          & 0.614          & 1.06                              & 0.764                               & 16.9                    \\
          FFNet       \cite{zhuSegmentationCoronaryArteries2022}    & 0.775          & 2.529          & 3.14                              & 0.785                               & 22.3                    \\
          3D-FFR-Unet \cite{songAutomaticCoronaryArtery2022}        & 0.859          & 0.262          & 0.53                              & 0.785                               & 39.7                    \\
          PSP-Net*    \cite{zhaoPyramidSceneParsing2017}            & 0.841          & -              & 0.59                              & -                                   & -                       \\
          GCB-Net*    \cite{zhaoGraphConvolutionBased2022}          & 0.899          & -              & 0.34                              & -                                   & -                       \\ 
          HMSA*       \cite{taoHierarchicalMultiScaleAttention2020} & 0.862          & -              & 0.56                              & -                                   & -                       \\
          DVS*        \cite{shinDeepVesselSegmentation2019}         & 0.873          & -              & 0.58                              & -                                   & -                       \\
          DDT*        \cite{wangDeepDistanceTransform2020}          & 0.882          & -              & 0.57                              & -                                   & -                       \\  \midrule
          \textbf{Ours}                                             & \textbf{0.895} & \textbf{0.193} & \textbf{0.38}                     & \textbf{0.054}                      & \textbf{2}                       \\ \bottomrule
      \end{tabular}        
  }
  \caption{Quantitative Evaluation Results of the Coronary Artery Segmentation for Different Methods on ASOCA Dataset. * denotes the results are quoted without the source code and more detailed metrics.}
  \label{table:asoca}  
\end{table}

To further verify the generalizability and robustness of our model, comparison experiments are carried out on the public ASOCA dataset, where the structures of the coronary artery are simple and clear. we follow the baseline provided by GCB-Net \cite{zhaoGraphConvolutionBased2022}. The results are shown in Table. \ref{table:asoca}, evidencing the feasibility and robustness of our method with the Dice of 0.895, HD of 0.193, ASSD of 0.38, Smooth of 0.054 and NoS of 2.

\begin{figure}[htbp]
  \centering
  \includegraphics[width=0.48\textwidth]{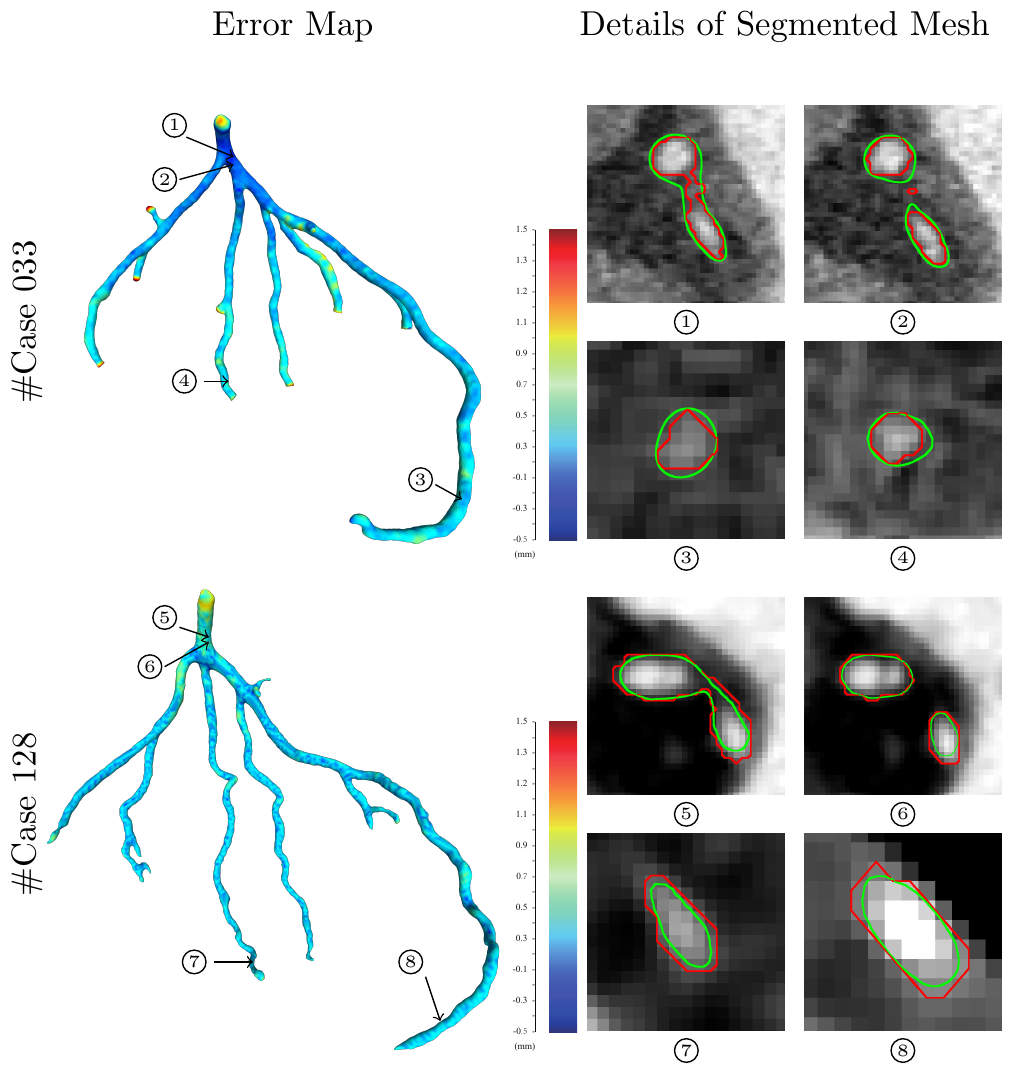}
  \caption{\textbf{Left.} Error maps between our results and ground truth. \textbf{Right.} Segmentation details of our methods. The red denotes the annotation labelled by radiologists, and the green represents our coronary artery segmented mesh results.}
  \label{fig:details}
\end{figure}

What's more, the error map between the generated coronary artery mesh and the ground truth is calculated to further illustrate our algorithm, and the detailed results of our geometry-based segmentation network and radiologists' annotations in CCTA images are exhibited as shown in Fig.\ref{fig:details}. In CCTA images, the red denotes the radiologists' annotations and the green represents the coronary artery mesh predicted by our model. As shown in the error map, the overall difference between the predicted mesh and the ground truth is particularly small, from -0.5 mm to 1.5 mm. The morphology of the coronary artery mesh generated by our geometry-based segmentation network is closely similar to the radiologists' annotations. From the details of the coronary artery shown in CCTA images, the generated coronary artery mesh has a naturally continuous transition at the multi-forks. Besides, the branches of the coronary artery mesh have a smooth, rounded and tubular structure, particularly at the ends with only a few discrete voxels such as \ding{174} and \ding{175} in Fig.\ref{fig:details}. 

\section{Conclusion}

In this paper, aiming at the complicated structures of the coronary artery with tiny and narrow branches, we propose a novel geometry-based segmentation network. With the assistance of the regularized mesh annotation, our model is competent for generating complete, smooth and elaborate results of the coronary artery, without the fragmentations of vessels. Extensive experiments verify our model, including our collected dataset CCA-200 and ASOCA, which show excellent quantitative results.

\bibliographystyle{unsrt}  
\bibliography{references}

\end{document}